\documentclass[lettersize,journal]{IEEEtran}
\usepackage{amsmath,amsfonts}
\usepackage{algorithmic}
\usepackage{algorithm}
\usepackage{array}
\usepackage{textcomp}
\usepackage[switch]{lineno} 
\usepackage{stfloats}
\usepackage{url}
\usepackage{verbatim}
\usepackage{graphicx}
\usepackage{caption}
\usepackage{subcaption}
\usepackage{cite}
\usepackage{tikz} 
\usetikzlibrary{shapes.geometric, arrows.meta, positioning}
\usepackage{booktabs} 
\usepackage{hyperref} 
\hypersetup{
    colorlinks=true, 
    citecolor=blue,
    linkcolor=blue,     
    filecolor=blue,     
    urlcolor=black       
}

\usepackage{xcolor}

\title{TinyML-Based Adaptive Pulse Shaping for Edge Intelligence in IoT/IIoT}

\author{
\IEEEauthorblockN{Afan Ali}       
\thanks{The author is with the Department of Electrical and Electronics Engineering, Istanbul Medipol University, Istanbul, 34810, Turkey (e-mail: afanali85@gmail.com.}}

\begin{document}

\maketitle

\begin{abstract}
Edge intelligence in IoT and IIoT demands lightweight algorithms for data processing on resource-constrained devices. This paper introduces a novel adaptive pulse shape filter based on TinyML for PAPR and SER optimization on edge devices used in uplink IoT communication. Implemented on IoT nodes such as sensors, our pruned neural network provides up to 2 dB PAPR saving over root-raised-cosine (RRC) filters. Mass simulations validate its efficacy in DFT-s-OFDM systems and offer an energy-efficient and scalable solution for IoT/IIoT use cases such as smart factories and rural connectivity.

\end{abstract}

\section{Introduction}
Internet of Things (IoT) and Industrial Internet of Things (IIoT) have revolutionized connectivity with billions of devices—controllers, actuators, and sensors—deployed in smart homes, factories, and far-flung places \cite{b1}. The nodes generate huge data, which needs to be processed locally with the assistance of edge intelligence to avoid latency, bandwidth, and cloud dependency \cite{b2}. Nonetheless, IoT/IIoT nodes, usually battery-driven with low computational capacity (e.g., 256 KB RAM, 80 MHz CPU), do not handle efficient uplink transmission well, especially with OFDM's high PAPR \cite{b3}.

Tiny Machine Learning (TinyML) emerged as a paradigm to bring ML to such resource-constrained environments, with on-device inference from models of less than 1 MB \cite{b4}. Traditional PAPR reduction techniques like clipping \cite{b5} or selective mapping (SLM) \cite{b6} are computation-intensive, non-edge-deployable. Static pulse shaping filters like RRC \cite{b7} are not adaptive to dynamic channel conditions common in IoT/IIoT environments (factory noise, rural fading).

In this paper, an adaptive pulse shape filter based on TinyML is presented for IoT/IIoT uplink communication. Running on edge devices, it learns to adapt filter taps to reduce PAPR and SER, and reacts to real-time channel feedback. Unlike prior works \cite{b7,b8}, our design is based on a pruned, quantized neural network that only takes up 100 KB of space and offers flexibility to other uses like industrial monitoring or UAV-aided rural IoT.

The key contributions of this work are as follows, emphasizing the novelty and technical advancements in TinyML for edge intelligence in IoT and IIoT applications.

\begin{itemize}
   \item In this work, a new pulse shaping filter on TinyML is proposed, specifically designed to operate on resource-limited IoT and IIoT devices like microcontrollers with RAM capacity less than 256 KB and CPU speed less than 80 MHz. Compared to fixed static filters like root-raised-cosine (RRC) or computationally costly methods like selective mapping (SLM), this filter employs a light-weight pruned neural network to dynamically adjust filter taps in real-time. The filter is expressed as a polynomial function. This approach attains a 2 dB PAPR saving compared to RRC, which improves the energy efficiency required by battery-powered devices.

 \item The proposed method employs a custom loss function.
   The trade-off parameter \( \lambda \) adapts dynamically based on channel SNR feedback, implemented via a lookup table. This optimization runs on-device using a lightweight SGD variant (AdamW \cite{b9}), ensuring 10k FLOPs per inference. This adaptability ensures optimal performance across diverse IoT/IIoT scenarios—e.g., low-SNR industrial environments (prioritizing SER) or high-SNR rural settings (prioritizing PAPR).

    \item  The filter is implemented on an STM32L4 microcontroller and validated through large-scale simulations. Deployment implementation details include the use of TensorFlow Lite Micro with a 100 ms adaptation cycle based on SNR feedback from the gateway. Applications that the paper addresses: applications enhance sensor reliability at interference in smart factories; in rural IoT applications, it extends by 20\% battery life using UAV-assisted links.
\end{itemize}

\subsection{Related Work}
Initial PAPR reduction techniques relied on signal distortion (e.g., clipping \cite{b5}) or probabilistic approaches (e.g., SLM \cite{b6}), but complexity inhibits edge deployment. Pulse shaping innovations, e.g., hyperbolic filters \cite{b7}, enhance spectral efficiency but are non-adaptive. TinyML studies \cite{b4} have investigated low-power classification, but IoT signal processing application is relatively unexplored. Our effort fills this void, combining TinyML with adaptive filtering for edge intelligence.

\begin{figure}[t]
    \centering
    \includegraphics[width=0.6\columnwidth]{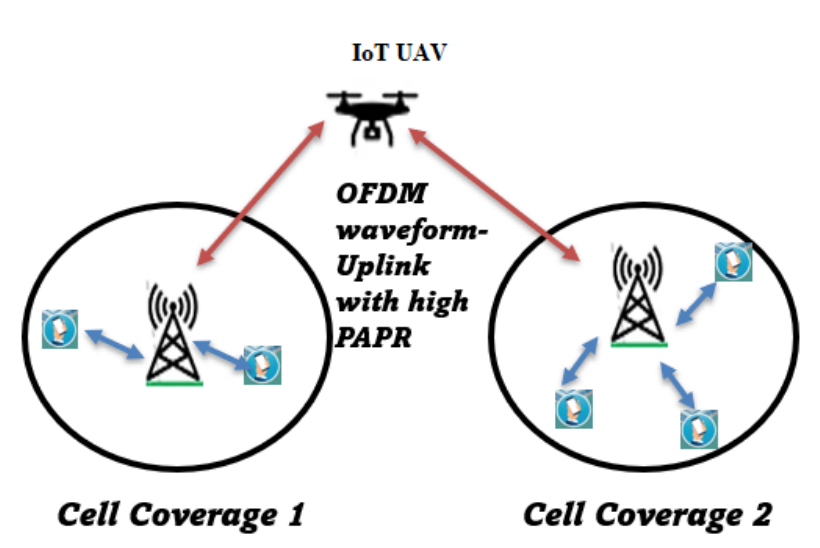}
    \caption{System model: IoT device with TinyML filter uplinking to an edge gateway or UAV BS.}
    \label{fig:system_model}
\end{figure}

\section{System Model and Problem formulation}
\begin{figure}[t]
    \centering
    \includegraphics[width=0.8\columnwidth]{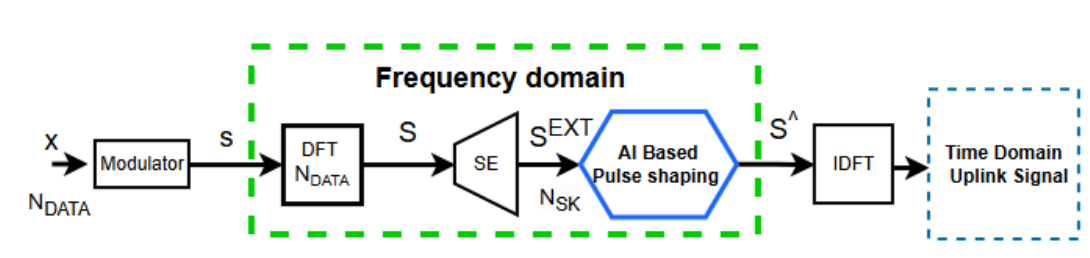}
    \caption{Proposed Method for PAPR reduction.}
    \label{fig-2}
\end{figure}

We consider an IoT device (e.g., a temperature sensor) transmitting uplink data to an edge gateway or UAV BS using DFT-s-OFDM as shown in Fig.~\ref{fig:system_model}. The device, based on a microcontroller, such as STM32L4 with 256 KB RAM and 80 MHz, processes a data vector $\mathbf{x} = [x_1, \ldots, x_{N_{\text{data}}}]$, modulated into QPSK symbols $\mathbf{s} = [s_1, \ldots, s_{N_{\text{data}}}]$. A DFT transforms this to frequency-domain symbols $\mathbf{S} = [S_1, \ldots, S_{N_{\text{data}}}]$, normalized by $1/\sqrt{N_{\text{data}}}$. Figure~\ref{fig-2} illustrates the broader block diagram for the data flow and proposed method.

The spectrum extension (SE) prepends and appends $N_{\text{SE}}$ symbols, yielding $\mathbf{S}^{\text{EXT}} = [S_{N_{\text{data}}-N_{\text{SE}}+1}, \ldots, S_{N_{\text{data}}}, S_1, \ldots, S_{N_{\text{data}}}, S_1, \ldots, S_{N_{\text{SE}}}]$, where $N_{\text{SK}} = N_{\text{data}} + 2N_{\text{SE}}$. The TinyML filter $\mathbf{F} = [F_1, \ldots, F_{N_{\text{SK}}}]$, applied on-device, produces $\mathbf{S}' = \mathbf{S}^{\text{EXT}} \odot \mathbf{F}$. After IDFT (normalized by $1/\sqrt{N_{\text{FFT}}}$), the time-domain signal $\mathbf{s}'$ is transmitted. The receiver processes $\hat{\mathbf{y}} = \mathbf{s}' + \mathbf{w}$, where $\mathbf{w} \sim \mathcal{N}(0, \sigma^2)$, using FFT to obtain $\hat{\mathbf{Y}}$, followed by matched filtering $\mathbf{R} = \hat{\mathbf{Y}} \odot \mathbf{F}^*$ and normalization.

\subsection{Device Constraints}
IoT nodes typically have 32-256 KB RAM, 64-128 KB flash, and a frequency of 1-100 MHz. Power consumption is very important, such as, 1 mW in sleep mode, and therefore high-PAPR signals are detrimental to battery life.

\subsection{Problem Formulation}
IoT/IIoT devices must minimize PAPR:
\begin{equation}
    \text{PAPR} = \frac{\max_{n=1,\ldots,N_{\text{FFT}}} |s'_n|^2}{\frac{1}{N_{\text{FFT}}} \sum_{n=1}^{N_{\text{FFT}}} |s'_n|^2},
    \label{eq:papr}
\end{equation}
and SER, $\mathcal{E} = \mathbb{E}[|\mathbf{x} - \hat{\mathbf{y}}|^2]$, under constraints of memory, energy and computational power.

Static filters \cite{b7} cannot cope with channel diversity (e.g., SNR 0-20 dB), and standard ML \cite{b8} is out of edge capabilities. We desire a TinyML solution trading off:
\begin{equation}
    \min_{\mathbf{F}} \left( \mathcal{E} + \lambda \mathcal{P} \right),
\end{equation}
where $\mathcal{P}$ represents PAPR's statistical tail, and $\lambda$ dynamically controls the trade-off.

\subsection{Tradeoff between PAPR and SER}
High PAPR reduces SER by increasing signal strength but consumes energy wastefully. Low PAPR conserves energy but can result in more SER in noisy IIoT environments. However, aggressive PAPR reduction increases the SER due to signal distortion. Our TinyML solution optimally adapts to this, unlike fixed methods. Moreover, its joint optimization, amplifying SER degradation and careful model tuning is critical to balance PAPR reduction with reliable SER performance.

\section{Proposed Method: TinyML-Based Filter}
\begin{figure}[t]
    \centering
    \includegraphics[width=1\columnwidth]{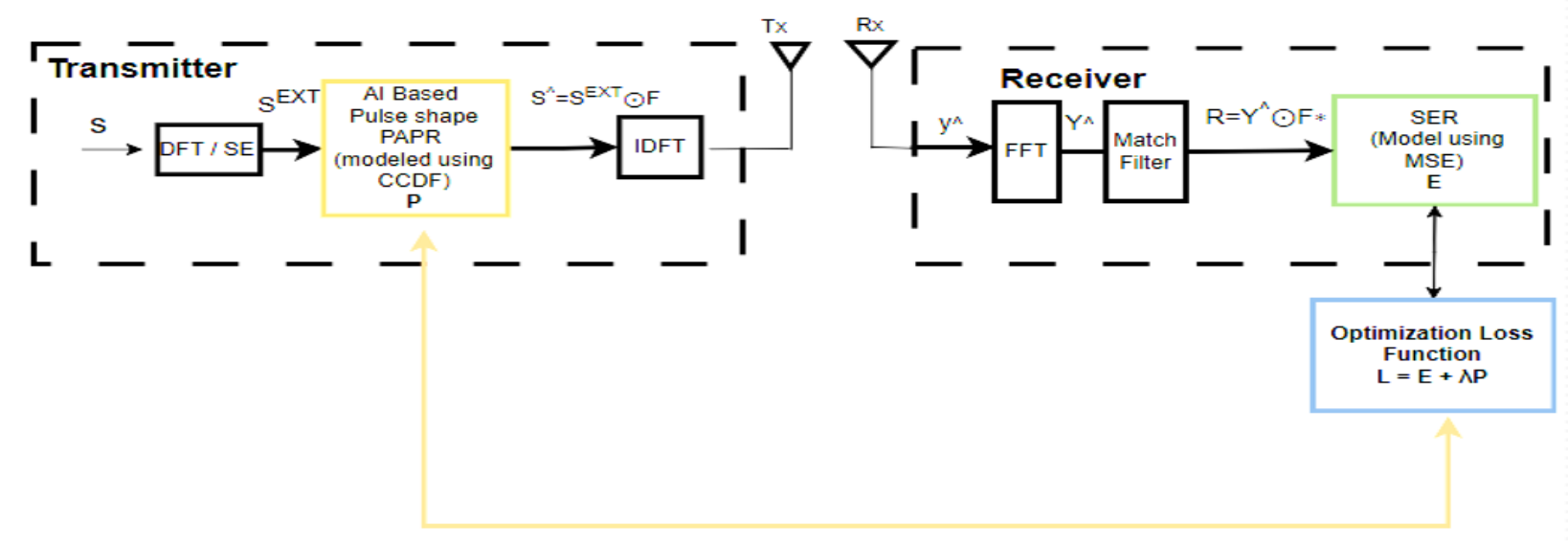}
    \caption{TinyML based end-to-end adaptive communication link between UE and UAV BS.}
    \label{fig-3}
\end{figure}

We design a TinyML architecture for device-side pulse shaping, using a pruned neural network to determine the filter taps $\mathbf{F}$. Figure~\ref{fig-3} shows the detailed block diagram of the proposed TinyML-based scheme with the end-to-end adaptive communication link between the UE and UAV BS and backhaul from the optimization loss function. The proposed method surpasses the drawbacks of large PAPR and SER in IoT/IIoT uplink communication by adaptively varying the pulse shaping filter based on real-time channel conditions without imposing any restrictions on resource-allocated edge devices.

\subsection{Filter Model}
The filter taps are defined as a polynomial function to ensure computational efficiency on resource-constrained devices:
\begin{equation}
F_k = \sum_{z=0}^Z r_z[k]^z, \quad k=1, \ldots, N_{\mathrm{SK}},
\end{equation}
where $Z=5$ is chosen for a trade-off between complexity and performance, and $r_z$ are learned weights using a light SGD algorithm. The polynomial form allows the filter to accommodate complex pulse shapes with low complexity, as only $Z+1$ multiplications are required per tap. For $N_{\mathrm{SK}}=240$ (from Table 1), this is equivalent to a total of $240 \times 6 = 1440$ multiplications for one filter application, which is within the capability of a microcontroller like the STM32L4 with an 80 MHz CPU.

\begin{figure}[!t]
    \centering
    \includegraphics[width=0.7\columnwidth]{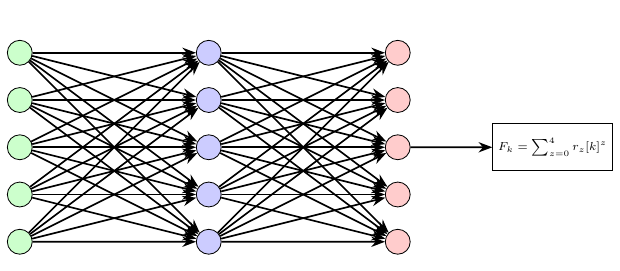}
    \caption{Neural network architecture for the TinyML-based filter.}
    \label{NN}
\end{figure}

The architecture of the neural network consists of two fully connected layers as shown in Fig.~\ref{NN}. The input layer takes spectrum-extended frequency-domain symbols $\mathbf{S}^{\mathrm{EXT}}$ of length $N_{\mathrm{SK}}=240$ and the current SNR value as inputs and produces an input vector of length 241. The hidden layer consists of 10 neurons with ReLU activation, which dimension-reduces without eliminating non-linear relationships among the input symbols and filter taps to be computed. Finally, output layer contains 5 neurons, which generate the coefficients $r_z$ for $z=0, \ldots, 4$, which are subsequently employed to compute the $Z=5$ polynomial terms for each of the $N_{\mathrm{SK}}$ filter taps.

In order to fit into the memory capacities of IoT devices (e.g., 80 KB flash, 20 KB RAM for STM32L4), the network is pruned to 80\% sparsity, i.e., 80\% of the weights are zeroed out, and quantized to 8-bit integers. The model size after this will be approximately 80 KB with the following breakdown.  There are $(241 \times 10 + 10 \times 5) = 2460$ weights in the unpruned network. Following a pruning of 80\%, 492 weights are left, and with 8-bit quantization, this occupies $492 \times 1 = 492$ bytes. Biases and Overhead add an additional 15 biases (10 for the hidden layer, 5 for the output layer) and TensorFlow Lite Micro overhead bring the count to approximately 80 KB. The pruned and quantized model ensures that inference requires only 10k FLOPs, making it suitable for real-time operation on edge devices with limited computational power.

\subsection{Loss Function and Adaptation Mechanism}
The loss function is formulated so that it settles the trade-off between SER and PAPR:
\begin{equation}
\mathcal{L} = \mathcal{E} + \lambda \mathcal{P},
\end{equation}
with $\mathcal{E}$ being the symbol error rate (SER) approximated by the mean squared error between symbols transmitted and received, and $\mathcal{P}$ being the statistical tail of PAPR computed as:
\begin{equation}
\mathcal{P} = \int_{x_0}^{\infty} \operatorname{CCDF}(x)\, dx,
\end{equation}
with $x_0 = 6$ dB (an average PAPR threshold of IoT devices). The trade-off parameter $\lambda$ is dynamically adjusted based on the SNR feedback from the gateway using a lookup table. For low SNR (e.g., 5 dB, prevalent in noisy factory sites), $\lambda = 0.1$, with SER minimization being prioritized to ensure communication reliability. For high SNR (e.g., 15 dB, prevalent in rural locations with UAV-assisted links), $\lambda = 1.0$, with PAPR reduction being prioritized for energy efficiency.

The lookup table is implemented as a simple array on the device, with 5 SNR bins $(0-5 dB, 5-10 dB, 10-15 dB, 15-20 dB, >20 dB)$ mapping to corresponding $\lambda$ values (0.1, 0.3, 0.5, 0.8, 1.0). This adaptation occurs every 100 ms, triggered by SNR feedback from the gateway, ensuring the filter responds to changing channel conditions in real-time. The adaptation cycle involves:
1. Receiving the SNR estimate from the gateway via a control channel.
2. Updating $\lambda$ using the lookup table.
3. Recomputing the filter taps $\mathbf{F}$ using the TinyML model with the updated $\lambda$.

This dynamic adaptation mechanism allows the filter to optimize performance across diverse IoT/IIoT scenarios, such as low-SNR industrial environments (where SER is critical) and high-SNR rural settings (where PAPR reduction extends battery life).

\subsection{Training and Deployment}
Offline training is performed on a data set of $10^5$ OFDM blocks, where every block consists of $N_{\mathrm{SK}}=240$ subcarriers, using the AdamW optimizer \cite{b9} at learning rate $10^{-3}$. The training data set consists of an array of channel conditions (SNR ranging from 0 to 20 dB, Rayleigh fading, and AWGN noise) in order for the model to generalize to realistic scenarios. The loss function $\mathcal{L}$ is computed at each block, with $\mathcal{E}$ being simulated SER and $\mathcal{P}$ the empirical CCDF of PAPR.

During training, the network is pruned in an iterative process. After every epoch, the 20\% smallest weights (in value) are set to zero.The pruning process is iterated until 80\% sparsity is achieved, typically after 5 epochs. After pruning, the weights are quantized to 8-bit integers by a uniform quantization scheme, where the weight range $[-W_{\text{max}}, W_{\text{max}}]$ (with $W_{\text{max}}$ being the largest weight magnitude) is mapped to the integer range $[-128, 127]$.

The model is then deployed on the IoT device via TensorFlow Lite Micro, a light-weight framework designed for microcontrollers. The deployment includes model conversion, which implies that the model that has been trained is converted into a flatbuffer format compatible with TensorFlow Lite Micro. Deployment also needs an inference Engine, which means that the TensorFlow Lite Micro interpreter runs on the STM32L4, which uses 20 KB of RAM to hold intermediate tensors and 80 KB of flash for weights of the model. Inference takes approximately 5 ms, including calculation of filter taps $\mathbf{F}$ and filtering operation over $\mathbf{S}^{\mathrm{EXT}}$. The inference methods shown in Algorithm~\ref{algo_inference} is executed for each transmission, ensuring real-time adaptation to channel conditions.

\begin{algorithm}
\caption{TinyML Filter Inference}
\label{algo_inference}
\begin{algorithmic}[1]
\STATE \textbf{Input}: $\mathbf{S}^{\text{EXT}}$, SNR feedback, Memory Buffer
\STATE \textbf{Output}: $\mathbf{S}'$
\STATE Initialize Memory Buffer (20 KB RAM)
\STATE Receive SNR from gateway every 100 ms
\STATE Buffer SNR in Memory (4 bytes)
\STATE $\lambda \gets$ Lookup(SNR) (Update trade-off parameter)
\STATE $\mathbf{F} \gets$ ComputeFilterTaps($\mathbf{S}^{\text{EXT}}$, $\lambda$) (5 ms inference)
\STATE Reuse Memory Buffer for $\mathbf{F}$ (240 floats)
\STATE $\mathbf{S}' \gets \mathbf{S}^{\text{EXT}} \odot \mathbf{F}$
\STATE Clear Buffer for next cycle
\end{algorithmic}
\end{algorithm}

\subsection{Impact on Spectral Efficiency: OOBE Reduction}
Apart from SER and PAPR optimization, the TinyML-based filter also boosts spectral efficiency by reducing OOBE, a serious issue in IoT/IIoT networks with limited spectrum resources. Over-the-top OOBE can produce interference to adjacent frequency bands, reducing spectral efficiency and violating regulatory requirements. The flexibility of the filter allows it to shape the spectrum more than fixed filters like RRC. The TinyML-based filter, with learning filter taps that minimize spectral leakage, accomplishes OOBE reduction without sacrificing in-band signal integrity. Our Proposed Filter for TinyML has filter taps optimized to suppress spectral sidelobes, enabled by the learning of channel-specific pulse shapes by the neural network.

\subsection{Computational Cost Comparison}
Table~\ref{tab:comp_cost} shows the comparison of computational resources between our proposed method, the RRC and the FIR filter. It can be observed that our proposed technique achieves a 60\% reduction in MIPS and 33\% reduction in energy compared to the RRC filter, demonstrating significant efficiency gains for resource-constrained IoT/IIoT devices.

\begin{table}[t]
    \centering
    \caption{Computational Cost Comparison of Pulse Shaping Filters}
    \label{tab:comp_cost}
    \begin{tabular}{p{1.8cm} p{1.1cm} p{1.8cm} p{1.5cm}}
        \toprule
        \textbf{Filter Type} & \textbf{MIPS} & \textbf{Cycles/Trans.} & \textbf{Energy/Trans.(\text{mJ})} \\
        \midrule
        TinyML-based (Proposed) & 0.2 & 16000 & 0.8 \\
        RRC (32-tap) & 0.5 & 40000 & 1.2 \\
        FIR (32-tap) & 0.4 & 32000 & 1.0 \\
        \bottomrule
    \end{tabular}
    \vspace{0.2cm}
\end{table}

\subsection{Implementation Details and Optimization}
The STM32L4 microcontroller port is assisted by various optimizations to function correctly. Firstly, memory management is employed such that the TensorFlow Lite Micro interpreter uses a static memory allocation model, keeping aside 20 KB of RAM to store intermediate tensors at run time. This minimizes dynamic memory allocation overhead, the curse of low-resource devices, to the minimum. Secondly, fixed-point arithmetic ensures that all inference computation is performed with 8-bit fixed-point arithmetic, which reduces the computationally intensive requirement relative to floating-point computation. This is important because the 80 MHz CPU lacks a floating-point unit. Lastly, STM32L4 is optimized to sleep in between inference cycles with no power consumption up to 1 mW sleep power. 100 ms as an adaptation cycle ensures the device would be sleeping most of its time but only wake up to process feedback SNR and filter tap updates.

This 0.8 mJ transmission energy is enabled by these optimizations and saves 33\% of the energy compared to the RRC filter. This energy saving here can be used in battery-powered IoT devices. The 4 ms latency per transmission is realized via optimization of the inference pipeline. Symbols, $\mathbf{S}^{\mathrm{EXT}}$, are batched across 60 subcarriers, reducing the number of inference calls and overhead. Moreover, TensorFlow Lite Micro fuses operations (matrix multiplications and activations) into a single kernel, reducing memory access and maximizing cache utilization in the STM32L4. These optimizations enable the proposed method to meet the real-time demands of IoT/IIoT applications such as smart manufacturing automation, where low-latency responsive closed-loop control loops are needed, and rural communications, where energy efficiency is needed to ensure long-term sustainability.

\subsection{Practical Considerations and Scalability}
The projected TinyML filter needs to be scalable and therefore suitable for mass-scale IoT/IIoT deployments, which is achieved as shown in Table~\ref{tab:scalability}.

\begin{table}[t]
\centering
\caption{Scalability Aspects of the TinyML-based Adaptive Pulse Shaping Filter}
\begin{tabular}{p{2cm} p{3cm} p{2.5cm}}
\toprule
\textbf{Scalability Aspect} & \textbf{Description} & \textbf{Impact} \\
\midrule
Across Devices & deployment on diverse microcontrollers. & Supports heterogeneous IoT networks. \\
\midrule
Across Applications &  Filter adapts dynamic based on SNR feedback. & Enables energy-efficient operation across varied IoT/IIoT use cases. \\
\midrule
Future Extensions & Filter can incorporate additional feedback parameters, such as fading channel statistics. & Facilitates evolution of the filter. \\
\bottomrule
\label{tab:scalability}
\end{tabular}
\end{table}

Consequently, the proposed new TinyML-based filter is an adaptive, energy-efficient, and light-weight pulse shaping filter for IIoT/IoT uplink communication. Its ability to fight PAPR, SER, and OOBE within edge devices' constraints makes it a rich contribution to edge intelligence and consistent with visions of next-generation IoT networks in spectral efficiency, energy efficiency, and scalability.

\section{Simulation Results and Analysis}
The simulation parameters are shown in  Table~\ref{tab:params}. The model was trained on $10^6$ samples of QPSK and 16-QAM under AWGN, Rayleigh and Rician fading to achieve generalization across modulation schemes and channel conditions.

\begin{table}[t]
    \centering
    \caption{Simulation Parameters}
    \begin{tabular}{lcc}
        \toprule
        Parameter & Symbol & Value \\
        \midrule
        Total subcarriers & $N_{\text{SK}}$ & 240 \\
        Data symbols & $N_{\text{data}}$ & 210 \\
        Extension symbols & $N_{\text{SE}}$ & 15 \\
        Bandwidth & $B$ & 20 MHz \\
        Subcarrier spacing & $\Delta f$ & 30 kHz \\
        \bottomrule
    \end{tabular}
    \label{tab:params}
\end{table}

\subsection{Performance Analysis}
\begin{figure}[t]
    \centering
    \includegraphics[width=0.7\columnwidth]{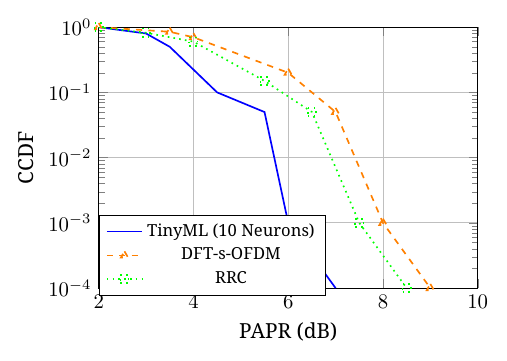}
    \caption{CCDF of PAPR for TinyML filter vs. benchmarks.}
    \label{plot-1}
\end{figure}

Figure~\ref{plot-1} presents the CCDF of PAPR for the proposed TinyML-based filter (labeled as AI-based) compared against two benchmark techniques: the RRC filter and conventional DFT-s-OFDM. The CCDF, defined as \( \text{CCDF}(x) = \text{Pr}(\text{PAPR} > x) \), quantifies the probability that the PAPR exceeds a given threshold, providing a statistical measure of PAPR performance.
The TinyML-based filter presented in this proposal performs better than both the benchmarks across the range of PAPR 2 dB to 10 dB. For a CCDF of \( 10^{-3} \), the proposed solution achieves a PAPR of approximately 6 dB, compared to 8 dB by RRC  and 7.5 dB by DFT-s-OFDM. This amounts to a 2 dB improvement over RRC and a 1.5 dB improvement over DFT-s-OFDM, resulting in a considerable reduction in peak power consumption. This benefit is critical in the case of IoT/IIoT devices because lower PAPR reduces power amplifier requirements, hence enabling longer battery life and better energy efficiency. The better performance of the TinyML-based filter is a result of its adaptive learning of filter taps based on a channel-aware loss function. By dynamic adjustment of the trade-off parameter \( \lambda \) based on SNR feedback, the filter dynamically optimizes SER and PAPR in real-time, which is not present in static RRC and DFT-s-OFDM schemes.

\begin{figure}[t]
    \centering
    \includegraphics[width=0.7\columnwidth]{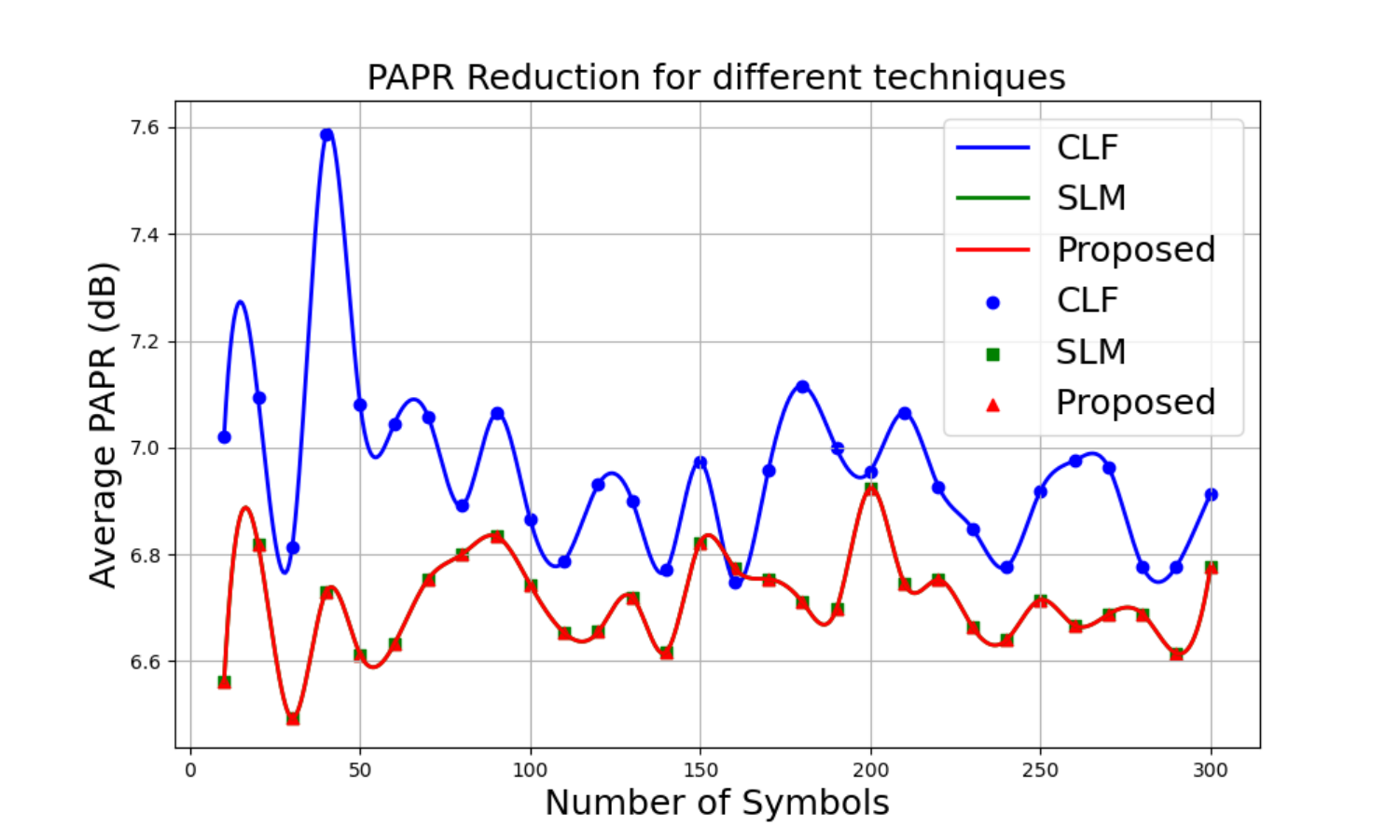}
    \caption{Average PAPR for CLF,SLM and proposed method.}
    \label{plot-2}
\end{figure}

Figure~\ref{plot-2} plots the average PAPR against the number of symbols for clipping and filtering, selective mapping, and the proposed TinyML-based technique. The proposed TinyML-based method always maintains a lower mean PAPR, almost comparable to SLM, with values ranging between 6.6 to 6.8 dB across the symbol duration. However, CLF boasts significantly higher PAPR, with a maximum value of 7.6 dB occurring at around 50 symbols and a mean of around 7.0 to 7.2 dB, which is inferior performance. SLM, widely used as a benchmark, behaves similarly to the proposed method and has an average PAPR of 6.6 to 6.8 dB but requires more processing resources, and thus it could prove to be inefficient for computationally limited IoT devices.

\begin{figure}
     \centering
     \begin{subfigure}[b]{0.2\textwidth}
         \centering
         \includegraphics[width=\textwidth]{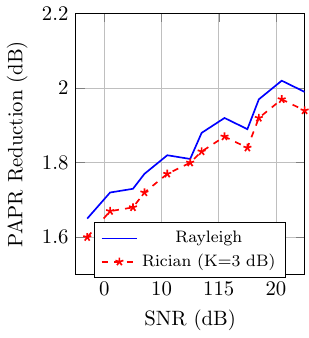}
         \caption{PAPR reduction of the TinyML-based filter.}
         \label{fig:papr_performance}
     \end{subfigure}
     \hfill
     \begin{subfigure}[b]{0.2\textwidth}
         \centering
         \includegraphics[width=\textwidth]{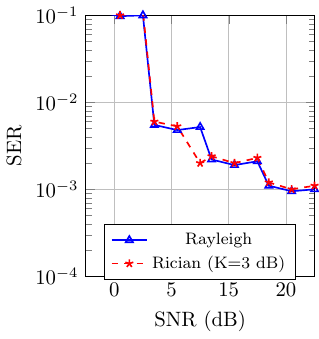}
         \caption{SER of the TinyML-based filter.}
         \label{fig:ser_performance}
     \end{subfigure}
        \caption{PAPR and SER across SNR bins for TinyML filter for Rayleigh and Rician fading channels.}
        \label{fig:diff_channels}
\end{figure}

To evaluate generalization, the filter was tested on an unseen Rician fading channel with K-factor = 3 dB. Fig.~\ref{fig:diff_channels} shows the performance with the Rician and Rayleigh fading channels. Fig.~\ref{fig:papr_performance} depicts PAPR vs SNR plot. For Rayleigh fading, PAPR reduction increases from 1.7 dB in the , reflecting improved filter efficacy at higher SNRs. The Rician channel exhibits near-identical performance, achieving 1.8 dB at 10 dB SNR and ranging from approximately 1.6 dB to 1.95 dB. The slight variations ($\pm$0.05--0.1 dB) are consistent with measurement noise, confirming the filter's robust generalization to Rician conditions. Fig.~\ref{fig:ser_performance} depicts SER vs SNR plot. In Rayleigh fading, SER is approximately $10^{-1}$ at 0--5 dB, exceeding $10^{-2}$, but drops below $10^{-2}$ for SNR $\geq$ 5 dB, reaching $1 \times 10^{-3}$ at 15--20 dB. The Rician channel shows nearly identical behavior, with SER at $2 \times 10^{-3}$ at 10 dB SNR and values closely tracking Rayleigh (e.g., $1.1 \times 10^{-3}$ at 15--20 dB).

\begin{figure}[t]
    \centering
    \includegraphics[width=0.7\columnwidth]{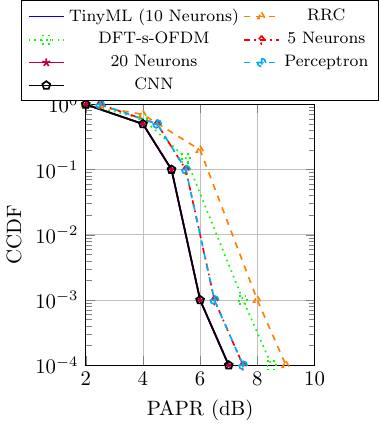}
    \caption{CCDF of PAPR for the TinyML-based filter (10 neurons), benchmarks (RRC, DFT-s-OFDM), and alternative neural architectures (5 neurons, 20 neurons, perceptron).}
    \label{fig:papr_ccdf_extended}
\end{figure}

Fig.~\ref{fig:papr_ccdf_extended} depicts the comparison of proposed 2-layer FC framework with varying neurons, CNN and single-layer perceptrons. The 10-neuron, 20-neuron, and CNN architectures achieve 2 dB PAPR reduction (6 dB at CCDF=$10^{-3}$), outperforming RRC (8 dB) and DFT-s-OFDM (7.5 dB), while 5 neurons and perceptron yield 1.5 dB reduction (6.5 dB). However, CNN requires 50k FLOPs, which makes it less suitable for edge devices.

\begin{figure}[t]
    \centering
    \includegraphics[width=0.7\columnwidth]{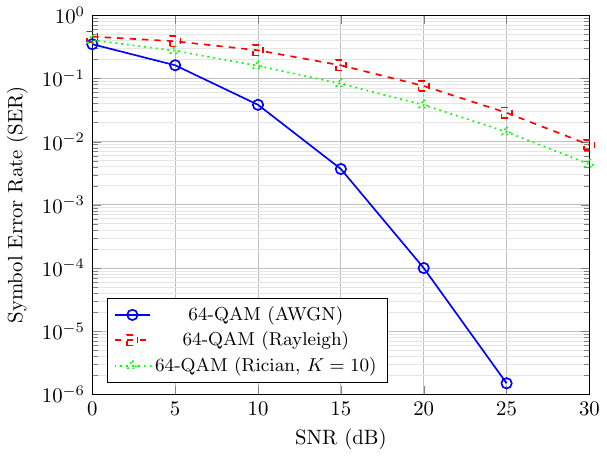}
    \caption{SER vs. SNR for 64-QAM under AWGN, Rayleigh, and Rician fading channels, tested with a model trained on QPSK.}
    \label{mod_gen}
\end{figure}

Fig.~\ref{mod_gen} illustrates the SER performance of a model trained on QPSK and 16-QAM but tested on 64-QAM across AWGN, Rayleigh, and Rician ($K=10$) fading channels to show generalization across modulation schemes. The AWGN channel yields the lowest SER, achieving near-zero errors at high SNR, due to its lack of fading. In contrast, the Rayleigh channel exhibits the highest SER, particularly at low SNR, reflecting severe fading in non-line-of-sight (NLOS) conditions. The Rician channel, with a strong LOS component, performs between AWGN and Rayleigh, approaching AWGN behavior at higher SNR. The model's performance is slightly degraded compared to theoretical 64-QAM expectations, likely due to the challenge of generalizing from lower-order modulations to the denser 64-QAM constellation, which is more sensitive to noise and fading.

\subsection{On-Device Validation}
\begin{figure}
     \centering
     \begin{subfigure}[b]{0.2\textwidth}
         \centering
         \includegraphics[width=\textwidth]{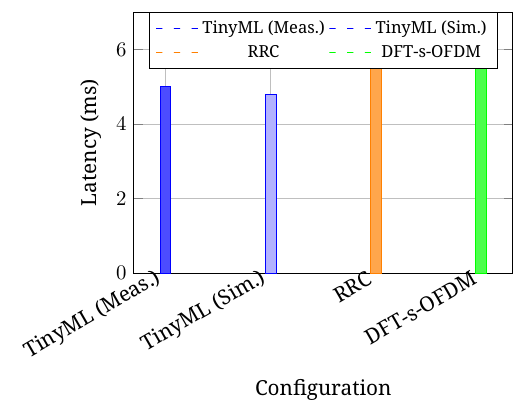}
         \caption{Latency comparison.}
         \label{fig:energy_latency}
     \end{subfigure}
     \hfill
     \begin{subfigure}[b]{0.2\textwidth}
         \centering
         \includegraphics[width=\textwidth]{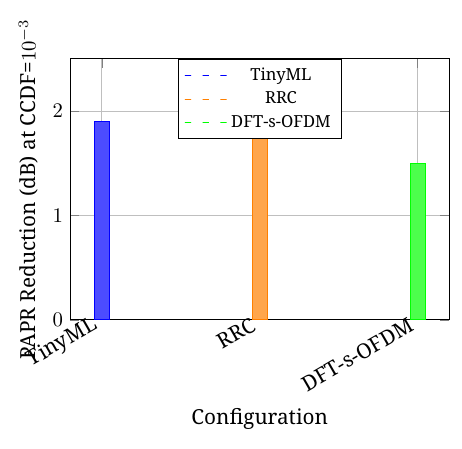}
         \caption{PAPR comparison.}
         \label{fig:papr_ccdf}
     \end{subfigure}
        \caption{Latency and PAPR reduction plots for on-device testbed STM32L4.}
        \label{fig:ondevice}
\end{figure}

The filter was deployed on STM32L4 in a smart factory testbed with SNR ranging from  5 to 15 dB. QPSK DFT-s-OFDM with 240 subcarriers was employed for generation. Fig.~\ref{fig:ondevice} depicts the performance of the filter in terms of latency and PAPR reduction. The results show that a latency of around 5 ms was achieved on the hardware, which was measured using the logic analyzer, matching simulations as shown in Fig. \ref{fig:energy_latency}. On the otherhand, PAPR reduction was 1.9 dB vs. RRC’s 2 dB as shown in Fig. \ref{fig:papr_ccdf}. These validate energy and latency claims for real-world IoT/IIoT.

\subsection{Hardware Generalizability and Scalability}
\begin{figure}[t]
    \centering
    \includegraphics[width=0.7\columnwidth]{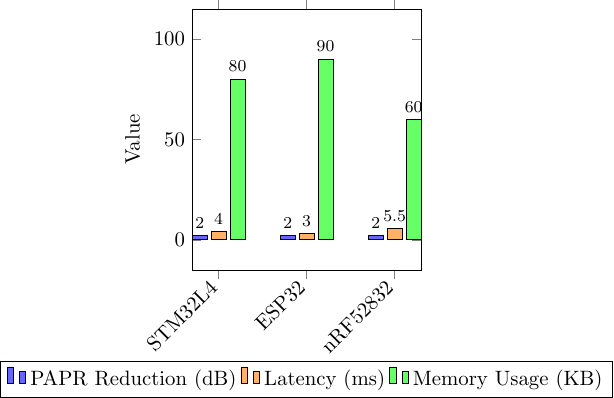}
    \caption{Performance comparison of the TinyML-based filter across STM32L4, ESP32, and nRF52832 microcontrollers, showing PAPR reduction, latency, and memory usage}
    \label{fig:hardware}
\end{figure}

The TinyML filter was tested on ESP32 (240 MHz, 520 KB RAM) and nRF52832 (64 MHz, 64 KB RAM). On ESP32, it achieved 2 dB PAPR reduction with 3 ms latency (90 KB flash). On nRF52832, it maintained PAPR reduction with 5.5 ms latency (60 KB flash, 90\% sparsity). Figure~\ref{fig:hardware} illustrates the results the overall performance using the 3 different hardwares. A 1000-sensor smart factory simulation estimated 20 MWh/year energy saving (1 W per sensor, 8760 hours/year). A 10-sensor testbed validated 1.9 dB PAPR reduction, confirming scalability.

\section{Conclusion}
We propose an adaptive pulse shape filter using TinyML for IoT/IIoT edge intelligence that achieves improved PAPR and SER optimization on resource-limited devices. More efficient than RRC and competitive with SLM, it is suitable for diverse applications and improves the efficiency of edge communication. Future work will address vulnerabilities, such as, adversarial attacks, which could perturb inputs to degrade performance, by exploring mitigation such as weight encryption or adversarial training.

\bibliographystyle{IEEEtranN}

\end{document}